\documentclass[letter]{aa} 

\usepackage{graphicx}
\usepackage{txfonts}
\begin{document}

   \title{Weak line discovered by Voyager 1 in the  interstellar medium: Quasi-thermal noise produced by very few fast electrons}

   \subtitle{ }

   \author{N. Meyer-Vernet
          \inst{1}
          \and
          A. Lecacheux\inst{1}
          \and
          K. Issautier\inst{1}
      \and
      M. Moncuquet\inst{1}
  }

   \institute{LESIA, Observatoire de Paris, PSL Université, CNRS, Sorbonne Université, Université de Paris, 92195 Meudon, France\\
              \email{nicole.meyer@obspm.fr,alain.lecacheux@obspm.fr,karine.issautier@obspm.fr,michel.moncuquet@obspm.fr}
                                                }

   \date{Received January 3, 2022; accepted February 8, 2022}

 
  \abstract{
     A weak continuous line  has been recently discovered onboard Voyager 1 in the interstellar medium, whose origin raised two major questions. First, how can this line be produced by plasma quasi-thermal noise on the Voyager short antenna? Second, why does this line emerge at some distance from the heliopause? We provide a simple answer to these questions, which elucidates the origin of this line. First, a minute quantity of supra-thermal electrons, as generally present in plasmas -- whence the qualifier `quasi-thermal' -- can produce a small plasma frequency peak on a short antenna, of amplitude independent of the concentration of these electrons; furthermore, the  detection required long spectral averages, alleviating the smallness of the peak compared to the  background. We therefore attribute the observed line to  a minute proportion of fast electrons that contribute negligibly to the pressure. Second, we suggest that, up to some distance from the heliopause, the large compressive fluctuations ubiquitous in this region prevent the line to emerge from the statistical fluctuations of the receiver noise  because it is blurred out by the averaging required for detection, especially in the presence of short-wavelength density fluctuations. These results open up novel perspectives for interstellar missions, by showing that a minute proportion of fast electrons may be sufficient to measure the density  even with a relatively short antenna, because the quietness of the medium enables a large number of spectra to be averaged. 
 }
  
\titlerunning{Weak line  in the ISM: quasi-thermal noise with few fast electrons}
\authorrunning{Meyer-Vernet et al.}


   \maketitle
%

\section{Introduction}

   \citet{ock21} 
    (see also \citealt{bur21}) have recently discovered a weak continuous line close to the  local plasma frequency on the Voyager 1 Plasma Wave System (PWS) instrument  \citep{sca77} in the  interstellar medium. The origin of this line was not understood since a plasma quasi-thermal noise (QTN) origin  was deemed problematic.

   The plasma QTN  is produced by the quasi-thermal motion of  ambient plasma electrons, which induce electric fluctuations \citep{sit67} detected in situ by electric antennas. At frequencies below the plasma frequency $f_p$, the electrons are Debye shielded, so their thermal motion produces voltage pulses shorter than the inverse frequency of observation, yielding a spectral plateau. Above $f_p$, the quasi-thermal motions excite Langmuir waves, producing a spectral peak near $f_p$ as well as a power spectrum proportional to the electron pressure at high frequencies \citep{mey89}. Since the  discovery of this noise in the interplanetary medium  \citep{mey79,cou81},  calculations  of it have been extended to flat-top \citep{cha89} and Kappa velocity distributions  \citep{cha91, zou08, lec09}, to magnetised plasmas  \citep{sen82, mey93}, and to include ions \citep{iss99}. Spectroscopy of this noise has been used to measure in situ the electron density and temperature and the properties of supra-thermal electrons in the solar wind as well as in planetary and cometary environments on the spacecraft ISEE3-ICE, Ulysses, WIND, Cassini (\cite{mey17} and references therein), and more recently, Parker Solar Probe  \citep[PSP;][]{mon20}.

   Since the Langmuir wave phase speed exceeds  the electron thermal speed by a larger and larger margin as the frequency approaches $f_p$, whereas electric antennas are mainly sensitive to wavelengths of the order of their length, the peak appears only on antennas longer than the Debye length if the plasma is Maxwellian. However, because of the strong increase in  the wave phase speed close to $f_p$, the presence of supra-thermal electrons of similar speed produces a peak that still exists for antennas shorter than the Debye length, albeit with a smaller amplitude \citep{mey89, cha91}. Figure 1 shows such an example where the  line appears clearly on the 2 m antenna of the PSP  FIELDS instrument  \citep{bal16}, even when the Debye length  estimated via QTN spectroscopy (bottom panel) exceeds the equivalent antenna length. This line amounts to about $0.5 \times 10^{-14}$ V$^2$/Hz at the receiver ports (Fig. 1), which corresponds to about $6 \times 10^{-14}$ V$^2$/Hz at the antenna ports (considering the gain factor); it was interpreted as QTN produced by a few percent of supra-thermal electrons of temperature of the order of 100 eV  \citep{mon20}.
   
    \begin{figure*}
   	\centering
   	\includegraphics[width=16cm]{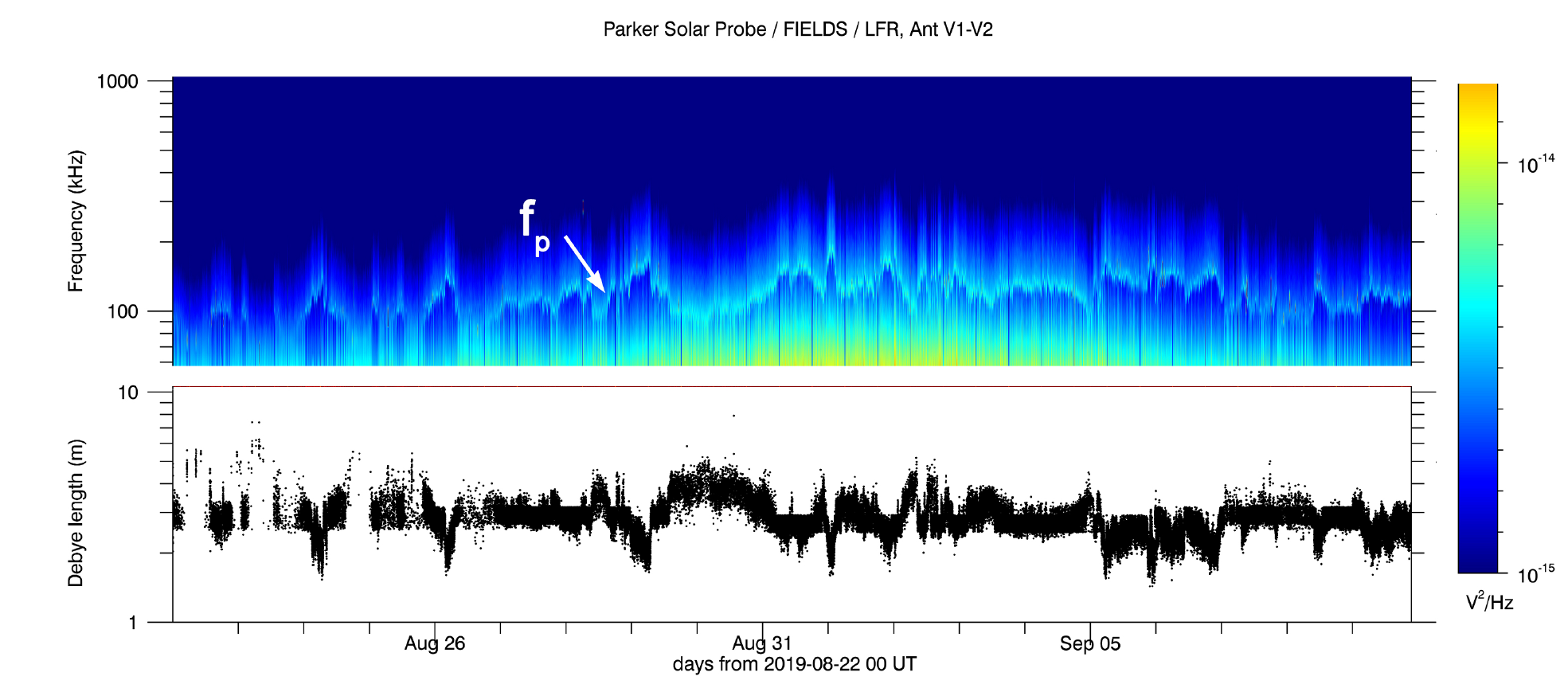}
     	\caption{Spectrogram acquired during the third PSP solar perihelion (August-September 2019) with the FIELDS antenna, showing the plasma QTN on which the $f_p$ line emerges clearly (cyan line). The plasma Debye length is plotted in black in the bottom panel.}
   	\label{Fig1}
   \end{figure*}

Therefore, a $f_p$ line can be observed in the presence of a small proportion of supra-thermal electrons, even with a relatively short antenna, if the radio receiver is sensitive enough (as is the case for the PSP/FIELDS instrument) or if the plasma frequency is sufficiently constant to enable it to emerge when a large number of spectra are averaged (as is the case of the  Voyager/PWS instrument, as we will show in the next section). In the interplanetary medium, such averaging cannot be performed because of the perturbations and the ubiquitous  short-wavelength density fluctuations \citep{cel87}. 

A crucial property of the peak is that its height only depends on the  velocity profile   at speeds close to the wave phase speed at the measuring frequency \citep{cha91, mey17}. In other words, the height of the peak depends on the energy of supra-thermal electrons, not on their concentration; in contrast, increasing this concentration tends to increase the peak  width. Hence, to explain the observed Voyager thin line, it is not necessary to assume 50\% supra-thermals, which increases the pressure by a large factor,  as  \cite{gur21} did.

\section{QTN in the quiet local interstellar medium
}

 The Langmuir wave number responsible for the line at frequency $f=f_p+\Delta f$ with $\Delta f/f_p \ll 1$ is
 \begin{equation}
  k_L = (\omega_p/v_{\mathrm{th}}) [2 \Delta f/f_p]^{1/2} \textrm{,} \label{kL}
 \end{equation}
 where $\omega_p = 2 \pi f_p$ and
 \begin{equation}
 v_{\mathrm{th}}^2 =  \langle v^2 \rangle = \int_0^\infty d^3v \; v^2 f(v) = 3 k_B T/m  \textrm{.} \label{T}
 \end{equation}
 Here $v_{th}$ is the electron mean square speed, $T$ the temperature, $f(v)$  the electron 3D-velocity distribution, and $m$ the electron mass. The phase speed producing the QTN near $ f_p$ is therefore
 \begin{equation}
 v_{\mathrm{ph}} \simeq  \omega_p/k_L = v_{\mathrm{th}} [ f_p/2\Delta f]^{1/2}  \textrm{.}  \label{vph}
 \end{equation}

\cite{ock21} indicate a frequency resolution of about $\delta f \simeq 10 $ Hz, of the same order of magnitude as the line width. This corresponds to $\Delta f/f_p  \simeq 3 \times 10^{-3}$, so Eq~(\ref{vph}) yields the wave phase speed $v_{\mathrm{ph}}  \simeq 7 \times 10^6 $ m/s at frequency $f_p + \delta f$, for $T=7000$ K. The electrons producing the QTN peak are thus expected to have an energy of the order of 100 eV or larger. We note that the relative velocity $v$ between the plasma and the spacecraft, which yields a relative Doppler-shift  $v/v_{\mathrm{ph}}\simeq 0.5 \%$, might modify the shape of the line with this frequency resolution.

The QTN peak at the  ports of an antenna of length $ L$ is given by \citep{mey17} 
\begin{equation}
V_f^2 \simeq \frac{8 m v_{\mathrm{ph}} F(\omega_pL/v_{\mathrm{ph}})}{\pi \epsilon_0  v_{\mathrm{th}}^2} \left[\frac{ \int _{v_{\mathrm{ph}}}^{\infty}  d v \; v \; f(v)}{ f(v_{\mathrm{ph}})} \right]  \textrm{,} \label{peak}
\end{equation}
where $v_{\mathrm{ph}} $ is given by (\ref{vph}) and  $F(x)$ is the antenna response, given for  $\omega_p L /v_{\mathrm{ph}} \ll 1$, by
\begin{equation}
F(x) \simeq x^2/24  \textrm{.} \label{F}
\end{equation}

We first  estimate the QTN peak by superimposing on the local interstellar medium (LISM)   velocity distribution (assumed to be a Maxwellian of density $n=0.11 $ cm$^{-3}$ and temperature $T = 7000$ K) a minute proportion $n_h/n \ll 1 $ of Maxwellian electrons of temperature $T_h \gg T$, namely $f_h \propto n_h T_h ^{-3/2} \exp (-mv^2/2k_B T_h)$. We see below that to  produce  a small enough line width,  $n_h$ should be so minute that it does not significantly change  the temperature $T$ (or the pressure).

A sufficiently large ratio $T_h/T$ can  produce a high peak, but for $L/L_D < 1 $ the peak width is set by the frequency at which the hot Maxwellian contributes to roughly half of the  distribution at the corresponding phase speed  \citep{mey89, mey17}, that is,
\begin{equation}
\exp(-mv_{\mathrm{ph}}^2/2k_B T) \simeq n_h T^{3/2} /(n T_h^{3/2})  \textrm{.} \label{width}
\end{equation}
Using Eqs~(\ref{T})-(\ref{vph}), we deduce the relative width of the peak as
\begin{equation}
\Delta f /f \simeq (3/4)/\ln[n T_h^{3/2}/(n_h T^{3/2})]  \textrm{.} \label{widthfin}
\end{equation}
Such a width,  depending in a logarithmic way on the concentration of supra-thermals,  is difficult to reconcile with the  very thin observed line, unless  an infinitesimal proportion of supra-thermal electrons is assumed.

Therefore, we instead  superimpose  on the   LISM  Maxwellian a minute proportion of electrons with  a power-law distribution $f_h (v) \propto 1/v^s$ (with $s > 2$) at speeds above  $ v_{\mathrm{min}} $, in a speed range that includes the phase speeds producing the line and a few times these speeds and with a density that ensures that the power law exceeds the Maxwellian at these speeds. Because $v_{\mathrm{ph}}/v_{\mathrm{th}}\gg 1$, the  Maxwellian has decreased by a huge factor at such speeds. Hence, the supra-thermal electrons largely dominate the total distribution at the speeds that produce the line, even if they contribute negligibly  to the electron density and temperature or pressure. We note that  the frequency resolution ensures that $v_{\mathrm{ph}} \ll c$ (the velocity of light in vacuum).

With such supra-thermal electrons, the whole distribution in Eq.~(\ref{peak}) can be replaced by $f(v) \propto 1/v^s$ at speeds $ v \ge v_{\mathrm{ph}} $, and so the bracket equals $ v_{\mathrm{ph}}^2 /(s-2)$. Therefore, Eqs~(\ref{vph})-(\ref{F})   yield at frequency $f_p+\Delta f$
\begin{equation}
V_f^2 =  \frac{2^{3/2}  \pi  m f_p^2 L^2}{3(s-2) \epsilon_0 v_{\mathrm{th}}}  \left(\frac{f_p}{\Delta f}\right)^{1/2} \label{peakfin}
\end{equation}
if $v_{\mathrm{ph}} \ge v_{\mathrm{min}}$, which implies from Eq~(\ref{vph})
\begin{equation}
\Delta f/f_p \le 3 k_B T/(2mv_{\mathrm{min}}^2)  \textrm{.} \label{width1}
\end{equation}

Approximating the peak level at frequency resolution $\delta f$ by the average,  $ \propto [\int_ {f_p} ^{f_p +\delta f} df/(f-f_p)^{1/2}]/\delta f = 2/\delta f^{1/2}$, we get at the antenna ports
\begin{equation}
V_f^2 \simeq  \frac{2^{5/2}  \pi  m f_p^2 L^2}{3(s-2) \epsilon_0 v_{\mathrm{th}}}  \left(\frac{f_p}{\delta f}\right)^{1/2} \label{peakfin1}
\end{equation}
if the relative frequency resolution $\delta f/f_p$ is smaller than the limit given by Eq. (\ref{width1})
; with $\delta f\simeq 10$ Hz, this requires a minimum energy of supra-thermals of about 100 eV. Level (\ref{peakfin1}) represents the QTN Langmuir wave contribution, to which, since $L<L_D$, one must add the remaining contribution of the order of the plateau level, which is roughly equal to  $10^{-15}$ V$^2$/Hz for these parameters \citep{mey89}. Since $k_L L\ll 1$, we substitute  $L=10/\sqrt2$ for the Voyager V antennas, as an approximation \citep{mey20}. Thus  Eq~(\ref{peakfin1}) yields $V_f^2 \simeq  10^{-14}/(s-2)$ V$^2$/Hz.

If $s=5-7$, which corresponds to a differential flux of  $dJ/dE \propto E^{-p}$ with $p = 1.5-2.5$ for sub-relativistic particles, we get $V_f^2 \simeq 0.3-0.4 \times 10^{-14}$ V$^2$/Hz. A roughly twice larger spectral resolution, as shown in Figs \ref{raie} and \ref{spdyn}, would change these values by a factor of $\sqrt{2}$, and the minimum energy of supra-thermals by a factor of 2. Given the approximations used, we estimate these levels to be accurate within a factor of 3.

\begin{figure}
	\centering
	\includegraphics[width=5cm]{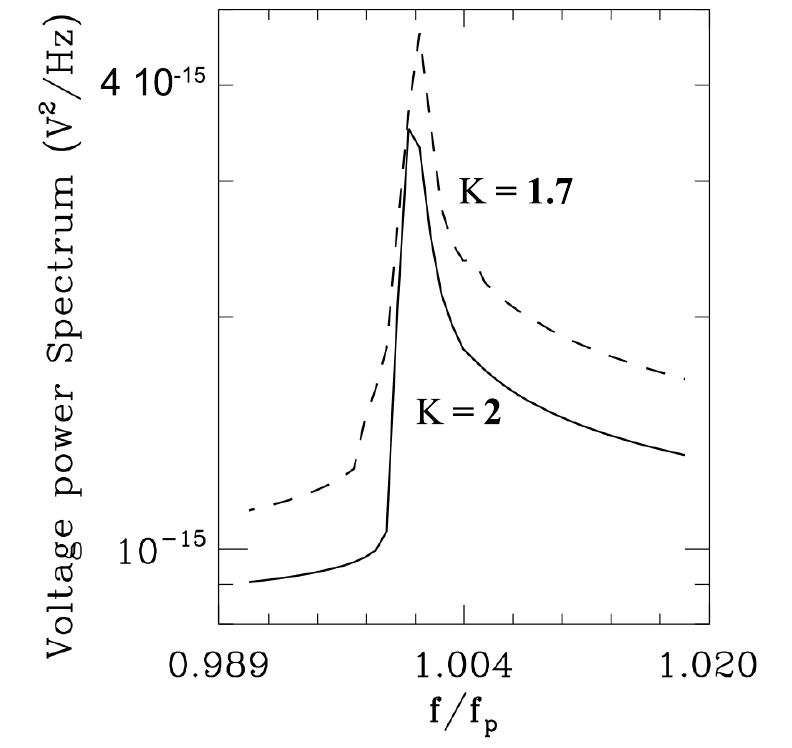}
	\caption{QTN $f_p$ peak produced by Kappa distributions, with  a relative frequency resolution $\delta f/f_p \simeq 10^{-3}$. As explained in the text, such distributions are convenient computational tools for estimating  the peak  produced by  superimposing on a Maxwellian a minute proportion of power-law-distributed electrons of speeds exceeding the wave phase speed. 
	}
	\label{Figspectre}
\end{figure}

Since  level (\ref{peakfin1}) is proportional to the square root of the inverse of the relative frequency  resolution, the line width will be of the same order of magnitude as the frequency resolution  if it is not widened by the density variations during the time of integration.

This is illustrated in Fig. \ref{Figspectre} which shows the $f_p$ line  calculated with a frequency resolution $\delta f/f \simeq 10^{-3}$ for Kappa distributions $\kappa = 2$ and $ 1.7$, respectively \citep{lec09}. Such distributions are convenient  for numerically estimating  the peak because at frequencies sufficiently close to  $f_p$, the phase speed $v_{\mathrm{ph}}$ is so large that (i) a minute proportion of Kappa-distributed supra-thermals   dominates the whole velocity distribution at $v_{\mathrm{ph}}$ and (ii) the Kappa itself reduces to a power law with $s=2(\kappa +1)$. Contrary to \cite{gur21}, we did not consider smaller values of $\kappa$, which would give rise to convergence problems, needing to be regularised \citep{sch18}.

Figure 2 also illustrates that, contrary to a common misunderstanding \citep{gur21}, the decrease in  Debye length as $\kappa$ decreases does not significantly increase the Langmuir wave contribution to the QTN (even though it changes somewhat the other contribution  that depends on low-energy electrons)  \citep{cha91}. This is because $k_L$ does not depend on $(1/\langle v^{-2} \rangle)^{1/2}$, which determines $L_D$, but on $v_{\mathrm{th}}= \langle v^2 \rangle^{1/2}$; only for Maxwellians are both quantities proportional.

The  measured amplitude cannot be calibrated because the gain values of the automatic gain control were not re-transmitted to the ground. However, the total average intensity can be compared  to that of  the background. Figure \ref{raie} shows such an example obtained when the line visibility is not ambiguous (i.e. after September 2017, see Fig. \ref{spdyn}) by averaging spectra over about  2600 seconds distributed over 2.5 years  in order to reduce the statistical fluctuations of the steady background level and make the line emerge.  One sees that the contribution of the line to the total intensity is smaller than the background by roughly one order of magnitude on average, and its   width is not significantly different from the achievable spectral resolution, given the expected broadening due to the uncertainty in frequency location and the expected intensity variations of the line over time. This weak line level confirms the results of \cite{ock21} and contradicts the interpretation by \cite{gur21} that the QTN should exceed the receiver noise.

\begin{figure}
	\centering
	\includegraphics[width=8.5cm]{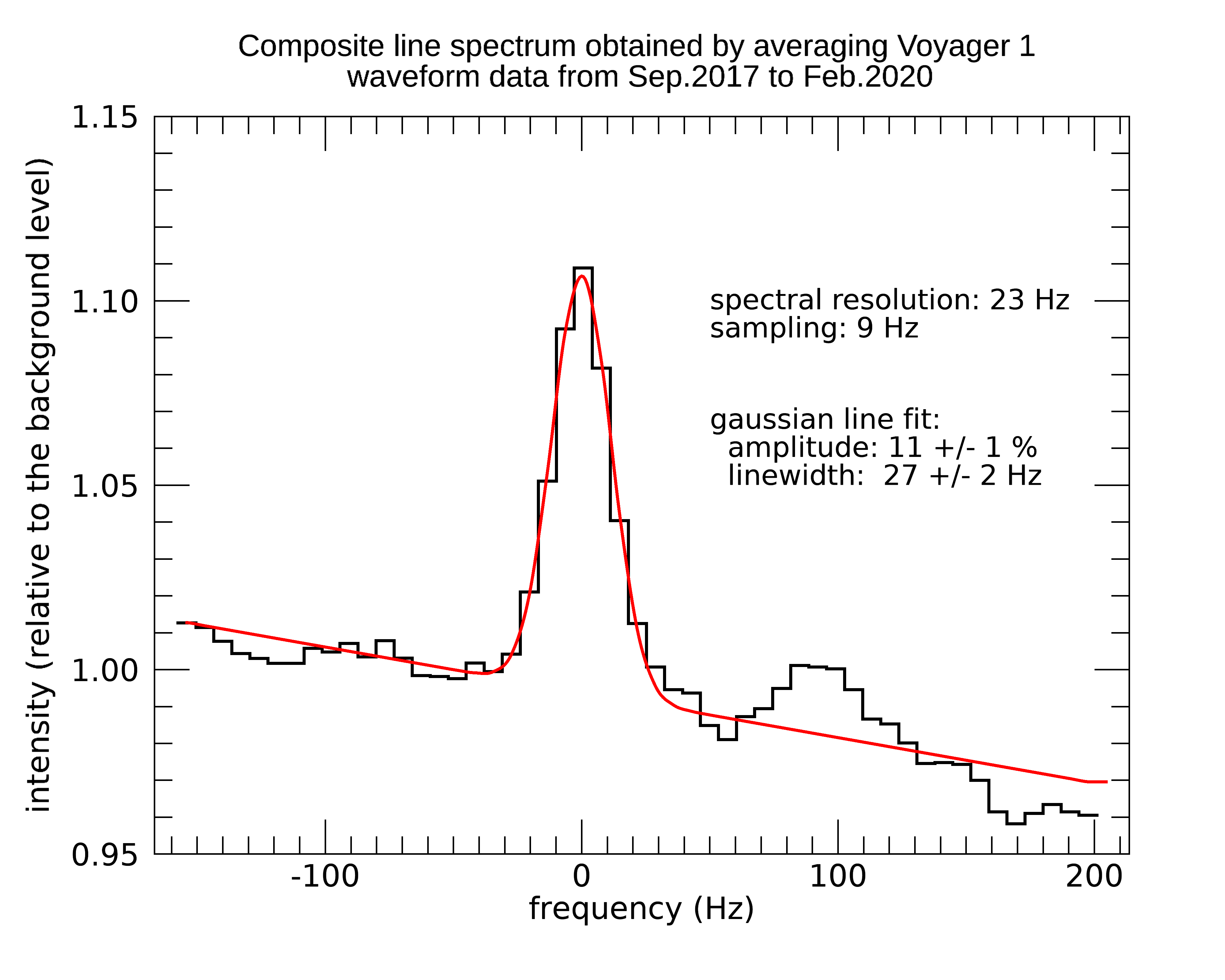}
	\caption{Line profile (linear scale) obtained by averaging 42600 spectra  shifted to a common central frequency by fitting a Gaussian profile (in red). The total intensity exceeds the background by 11\%, which means that the contribution of the line is smaller than the background level by about one order of magnitude  and that the  line width is not significantly different from the spectral resolution.
	}
	\label{raie}
\end{figure}

With the published receiver noise \citep{kur79, gur21}, this yields an average measured level  of the order of $ 10^{-14}$ V$^2$/Hz, which is marginally compatible with our estimates. However, the  receiver noise is very probably  overestimated below 20 kHz, since the published values are close to  the sum of the quasi-thermal and the shot noise at the beginning of  the Voyager mission. Hence the minimum measured level at these frequencies was not set by the receiver noise, but by the QTN, which represents a radio background  \citep{mey00}   unrecognised at this epoch.  Furthermore, the published receiver noise  was deduced by multiplying by the square of the  equivalent antenna length  for electromagnetic waves, which  has since been revaluated  to a smaller value \citep{mey96}.

Therefore,  our result is expected to explain the observed line, both in terms of amplitude and width.

\begin{figure*}
	\centering
		\includegraphics[width=16cm]{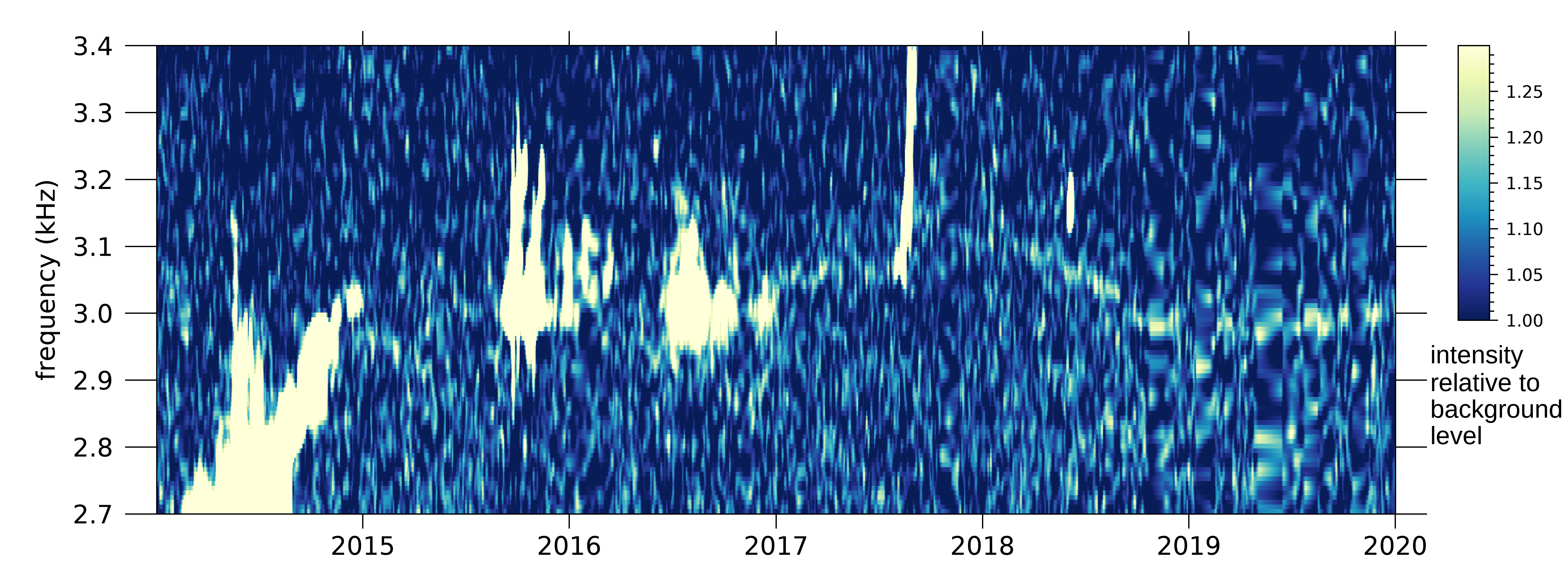}
	\caption{Spectrogram of the interstellar line, built from Voyager 1 PWS waveform data with a spectral resolution of 23 Hz (using Fourier transforms  of 1600 waveform samples, clocked at 28.8 kHz and Hamming-windowed).
	}
	\label{spdyn}
\end{figure*}

\section{Emergence of the  line at several AU from the heliopause
}
  
It has been suggested that the lack of observation of the line close to  the heliopause is due to the QTN decrease by the factor $\cos^2 \theta$ where  $\theta$ is the angle between the antenna effective axis and the magnetic field $\textbf{B}$ \citep{gur21}. However, this  interpretation   would require that the  electric field be aligned with $\textbf{B}$, which is doubtful because of the smallness of $B$ and because of the frequency resolution. Indeed,  the electric field is instead perpendicular to $\textbf{B}$ at the upper-hybrid frequency $f_{UH}=(f_p^2 + f_B^2)^{1/2}$, where  $f_B = eB/(2 \pi m)$ is the electron gyrofrequency  \citep{sti62}. Since $f_B/f_p = 3.7\times 10^{-3}$,  $f_{UH}$ differs from $f_p$  by less than $10^{-5}f_p$, which is much less than the frequency resolution. In other words, the QTN line is not expected to be affected by the magnetic field. This can also be understood from the dispersion equation of the generalised Langmuir mode \citep{wil00} $f^2(k, \theta) = f_p^2+f_B^2 \sin^2 \theta + (k v_{\mathrm{th}}/2 \pi)^2 $,    $\theta$ being the angle between  $\textbf{B}$ and the longitudinal electric field, since the $f_B$ term is negligible.

We suggest instead that the detection of the line is impeded when density fluctuations alter  its frequency, $f_p$, too much  during the time of integration required to detect the line. These fluctuations broaden the peak in relative value by one-half of the relative density fluctuations and decrease its height by roughly the ratio of the original peak width to the broadened value. This is the reason why one does not integrate the QTN noise over timescales exceeding a few seconds in the interplanetary medium, where perturbations and short-wavelength density fluctuations are ubiquitous \citep{cha91}.

Given the velocity of Voyager 1 (about 17 km/s in 2020) and PWS sampling occurring every two days at best, the medium was probed at points several millions of kilometres apart. In the quiet interstellar medium where the $f_p$ line is detected, the density fluctuations are so small that the frequency of the $f_p$ line changes very weakly, and is not blurred out after averaging. The value $\Delta n/n = 0.034$ found by \cite{ock21} was derived by integrating the fluctuation spectrum up to the larger observed fluctuation scale of 10 AU, which corresponds to 2.8 years. A sufficiently small time of integration  would broaden the line by less than the frequency resolution.  However, the density fluctuations are expected to be higher closer to the heliopause, where compressive fluctuations in the heliosheath are transmitted across the heliopause (Burlaga et al. 2015, 2018), but they are not expected to reach large distances farther out (Zank 2019). Depending on their amplitude and spectrum at scales corresponding to the time of integration, the line may be blurred out, especially if the short-wavelength turbulence, including the enhanced turbulence at kinetic scales \citep{lee19} is important. Conversely, the  disappearance of the line for a given time of integration might be used to deduce the amplitude of the density  fluctuations. We also note that the numerous solar-wind-initiated perturbations in this region producing intense  emissions (see Fig. \ref{spdyn}) impede  long-term averages, and the large-scale variations in density compromise the localisation of the line necessary to detect it. 
  
\section{Discussion and conclusion
}   
  
We have shown that superimposing a minute quantity of  electrons with a hard power-law velocity distribution above about 100 eV on the LISM Maxwellian at 7000 K  produces a QTN peak that explains the observed line. The width is of the order of the frequency resolution if the density fluctuations during the time of integration do not broaden the line, and the amplitude explains the observations, even better if the receiver noise level is a little lower than what has been published.

As in many dilute plasmas, the presence of supra-thermal electrons is not surprising  \citep{scu19}, since the Coulomb free path increases as the energy squared.  With an ambient electron density of the order of $n\simeq 0.1$ cm$^{-3}$, the Coulomb free path of   electrons of energy $E \simeq 100$ eV is about $l_{c(\mathrm{AU})} \simeq 0.05 E_{\mathrm{eV}}^2/n_{\mathrm{cm}^{-3}}\simeq 5000 $ AU. This is very large compared to the relevant scales and  to distances from shocks  that may produce such electrons and  to distances from  the heliopause. Several other equilibrium processes are operating \citep{dra18}, but they concern smaller energies.

Since the region close to the heliopause is filled with compressive fluctuations from the heliosheath that do not penetrate very far into the quiet interstellar medium, we suggest that the corresponding density fluctuations may blur out the  line there, especially if the short-wavelength turbulence is important, whereas these fluctuations have sufficiently decreased farther out to have only a small effect. Furthermore, the long-term variations in density may compromise the localisation of the line required in the detection process.  More detailed comparisons of the QTN theory with the observations with different averaging at different epochs could reveal more properties of the medium.

The present calculations do not consider the contribution of cosmic rays to the QTN,  which takes place  much closer to $f_p$ than the frequency resolution. However, it might be interesting to generalise the  calculations to relativistic speeds, in order to estimate this contribution.

Our results provide new perspectives for interstellar probes, showing for the first time that, thanks to the quietness of the medium, a minute proportion of supra-thermal electrons may be sufficient to enable the electron density to be measured with  antennas shorter than the Debye length, even though long antennas are mandatory  for accurately measuring the electron  temperature.

\end{document}